\documentclass[fleqn,usenatbib]{mnras}

\usepackage[T1]{fontenc}
\usepackage{ae,aecompl}
\usepackage{listings}
\usepackage[dvipsnames]{xcolor}

\usepackage{graphicx}	
\usepackage{newtxtext,newtxmath}
\usepackage{amsmath, amssymb, bm}	

\newcommand{\Nbody}{$N$-body }
\newcommand{\rj}{r_{\rm J}}
\newcommand{\rh}{r_{\rm h}}
\newcommand{\omegavec}{\boldsymbol{\omega}}
\newcommand{\trhi}{t_{\rm rh,i}}
\newcommand{\phiomega}{\phi_{\omega}}
\newcommand{\thetaomega}{\theta_{\omega}}

\title[Complex rotating star clusters]{The complex kinematics of rotating star clusters in a tidal field}

\author[M. Tiongco, E. Vesperini and A.~L. Varri]{
Maria A. Tiongco,$^{1}$\thanks{E-mail: mtiongco@indiana.edu}
Enrico Vesperini,$^{1}$
and Anna Lisa Varri$^{2}$
\\
$^{1}$Department of Astronomy, Indiana University, Bloomington, IN 47405, USA\\
$^{2}$Institute for Astronomy, University of Edinburgh, Royal Observatory, Blackford Hill, Edinburgh EH9 3HJ, UK
}

\date{Accepted XXX. Received YYY; in original form ZZZ}

\pubyear{2017}

\begin{document}
\label{firstpage}
\pagerange{\pageref{firstpage}--\pageref{lastpage}}
\maketitle

\begin{abstract}
We broaden the investigation of the dynamical properties of tidally perturbed, rotating star clusters by relaxing the traditional assumptions of coplanarity, alignment, and synchronicity between the internal and orbital angular velocity vector of their initial conditions. We show that the interplay between the internal evolution of these systems and their interaction with the external tidal field naturally leads to the development of a number of evolutionary features in their three-dimensional velocity space, including a precession and nutation of the global rotation axis and a variation of its orientation with the distance from the cluster centre. In some cases, such a radial variation may manifest itself as a counter-rotation of the outermost regions relative to the inner ones. The projected morphology of these systems is characterized by a non-monotonic ellipticity profile and, depending on the initial inclination of the rotation axis, it may also show a twisting of the projected isodensity contours.
These results provide guidance in the identification of  non-trivial features 
which may emerge in upcoming investigations of star cluster kinematics and a dynamical framework to understand some of the complexities already hinted by  recent observational studies.
\end{abstract}




\begin{figure*}
	\includegraphics[height=1.6in]{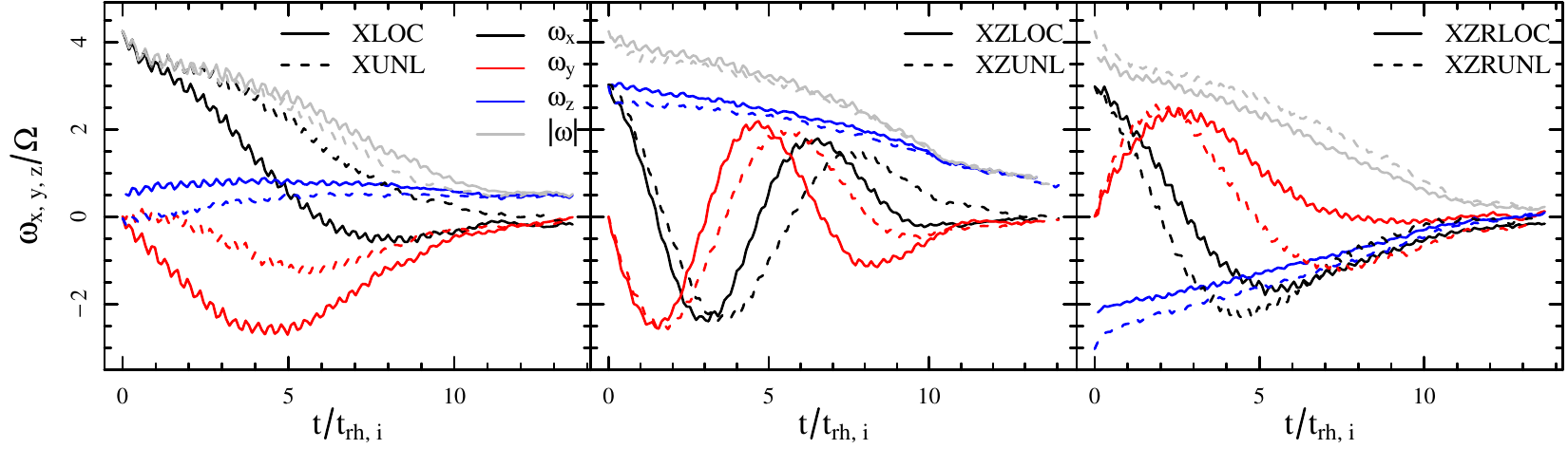}
    \caption{Time-evolution of the Cartesian components and the absolute value of the angular velocity vector, $\omegavec$, normalized to the angular speed of the cluster's orbit about the host galaxy, $\Omega$. All models are represented (see Table \ref{tab:table1}). Time is expressed in units of the initial half-mass relaxation time. The overall rotation of the systems decreases in magnitude (gray line), but in all cases, the tidal torque due to the host galaxy initiates a precession of the cluster's rotation axis leading to large scale oscillations of $\omega_{\rm x}$ and $\omega_{\rm y}$ (black and red lines).}
    \label{fig:omglob}
\end{figure*}

\section{Introduction}
Many recent observational studies are significantly enriching our knowledge of the internal structural and kinematical properties of globular clusters, and in particular revealing their deviations from the traditional picture of them being spherical and isotropic systems.  A consistent finding in these studies is the presence of rotation in  globular clusters (e.g., \citealt{bellazzini2012,fabricius2014,lardo2015}).

In most cases, the depth of current kinematic information allows only a simplified description of the internal distribution of the angular momentum. Ordered motions are often described only 
by a single value of the rotational velocity, and the orientation of the projected rotation axis is determined by using all members
in a given sample, often binning together stars that are at different distances from the centre of the cluster and thus have different kinematical properties. 
 A few studies have gone further and constructed a rotational radial profile of the cluster, (e.g., \citealt{gebhardt2000,vandenbosch2006,bianchini2013,boberg2017,bellini2017,kamann2017}), and some of these studies have noted that the rotation axis' position angle varies significantly with radius, suggesting that the orientation of such an axis may change with the distance from the cluster's centre. This complex kinematical property is usually not addressed in theoretical star cluster studies.

Indeed, only relatively few dynamical investigations have explored some fundamental aspects of the evolution of rotating clusters and shown that collisional systems gradually lose their rotation due to the effects two-body relaxation (e.g., \citealt{einsel1999,ernst2007,hong2013,tiongco2017}); this implies that present-day clusters had stronger rotation in the past and that rotation is an essential ingredient for more realistic models of cluster dynamics.  These studies have considered exclusively isolated 
or tidally limited rotating clusters with the rotation axis parallel or anti-parallel to the cluster's orbital angular velocity vector. In this contribution, we broaden the exploration of the dynamics of rotating clusters and focus our attention on more general 
initial configurations for which the internal rotation axis is not aligned with the cluster's axis of rotation about the host galaxy and the internal and orbital motions are not synchronous.

By means of \Nbody simulations, we explore the effects of the interplay between internal angular momentum and the 
external tidal field on the cluster's internal 
kinematics and show a dynamical path to a number of complex features in the 
three-dimensional velocity space of  star clusters, including the radial variation of the rotation axis hinted by some observational star cluster studies mentioned above. This investigation of the coupling between internal and orbital angular momenta may also help to interpret some kinematic features identified in more massive  stellar systems, such as nuclear star clusters (e.g., see \citealt{feldmeier2017}). Analogies to peculiar kinematic subsystems identified both in dwarf and massive early-type galaxies (e.g., see \citealt{franx1988} and  \citealt{derijcke2004}) may also be qualitatively drawn. 
\section{Method and Initial Conditions}

The $N$-body simulations presented in this paper were carried out using the {\sevensize NBODY6} code \citep{aarseth2003} accelerated by a GPU \citep{nitadori2012}, and run on the BIG RED II cluster at Indiana University. The clusters are scaled to \Nbody units such that $G=M=1$ and $E=-0.25$ \citep{heggie2003}.

For the initial conditions, we sample a series of rotating equilibria introduced by \citet{varri2012}.  
This family of distribution function-based models is characterized by differential rotation and an oblate structure flattened in a direction parallel to the rotation axis. The initial half-mass relaxation time, $\trhi$, is 62 times the half-mass orbital time ($\sqrt{3\pi/G\bar{\rho}}$).  For a complete description we refer the reader to \citet{varri2012}, and the parameters defining the specific models explored here are the same as those of the rotating models studied in \citet[see their \S 2]{tiongco2016a}.
In that study, the rotation axis of our models was assumed to be parallel to the angular velocity vector associated with the cluster's orbital motion.  In this 
Letter, we explore more general initial configurations in which the internal rotation axis is not aligned with  the angular velocity vector of the cluster's orbit about the host galaxy.

Clusters are assumed to be on circular orbits in the host galaxy tidal field modeled as a point-mass, and the equations of motion are solved in a frame of reference co-rotating with the cluster around the host galaxy centre (see e.g. \citealt{heggie2003}) with angular speed $\Omega$. 
When we present the results of our analysis in this paper, however, we use a non-rotating reference frame still 
centred on the cluster's centre with the $z$-axis perpendicular to the cluster's orbital plane.
The three-dimensional direction of the angular velocity vector, $\omegavec$, is expressed by $\theta_{\omega}$, i.e. the angle measured from the z-axis, up to 180 degrees, and $\phi_{\omega}$, which is defined 
as the angle between the projection of the vector on the $x-y$ plane and the $x$-axis, starting at 0 degrees at the $x$-axis and increasing towards the $y$-axis. 
We consider systems with $N=32~768$ equal-mass particles; particles moving beyond a distance from the cluster's centre equal to
two times the Jacobi radius, $\rj$, are removed from the simulation.  All the simulations are run until 75\% of the initial cluster mass is lost.

Table \ref{tab:table1} lists the simulations presented in this Letter and the initial orientation of the rotation axis of the cluster.  
All models have their initial rotation axis in the $x-z$ plane, therefore, only the $\thetaomega$ values of the initial orientation of the rotation axis are reported (see Table \ref{tab:table1}, Col. 2).
For each orientation we run two models: one including only the intrinsic rotation of the \citet{varri2012} model, and one in which we add also a positive solid body rotation about the $z$-axis with an angular speed, $\Omega$, equal to that of the cluster's orbital motion.  
This set-up is intended to generalize the usual initial condition in which a cluster model is assumed to be co-rotating with the orbital motion. We refer to such models as `locked', indicating the the solid-body component of the rotation has been assumed to come from tidal locking. For simplicity, the models without such a initial solid-body component are defined as `unlocked'.
We initialize the ratio of the half-mass radius of the cluster to the Jacobi radius to be $\rh/\rj$ = 0.17.

\begin{table}
\caption{Summary of simulations}
\label{tab:table1}
\begin{tabular}{@{}cccc}
\hline
Model ID & 
$\theta_{\rm rot}$ (degrees) &
Locked/Unlocked\\
\hline
XLOC &  90 & Locked \\
XUNL &  90 & Unlocked \\
XZLOC & 45 & Locked \\
XZUNL & 45 & Unlocked \\
XZRLOC & 135 & Locked \\
XZRUNL & 135 & Unlocked \\
\hline
\end{tabular}
\end{table}

\section{Results}

\begin{figure*}
	\includegraphics[height=1.6in]{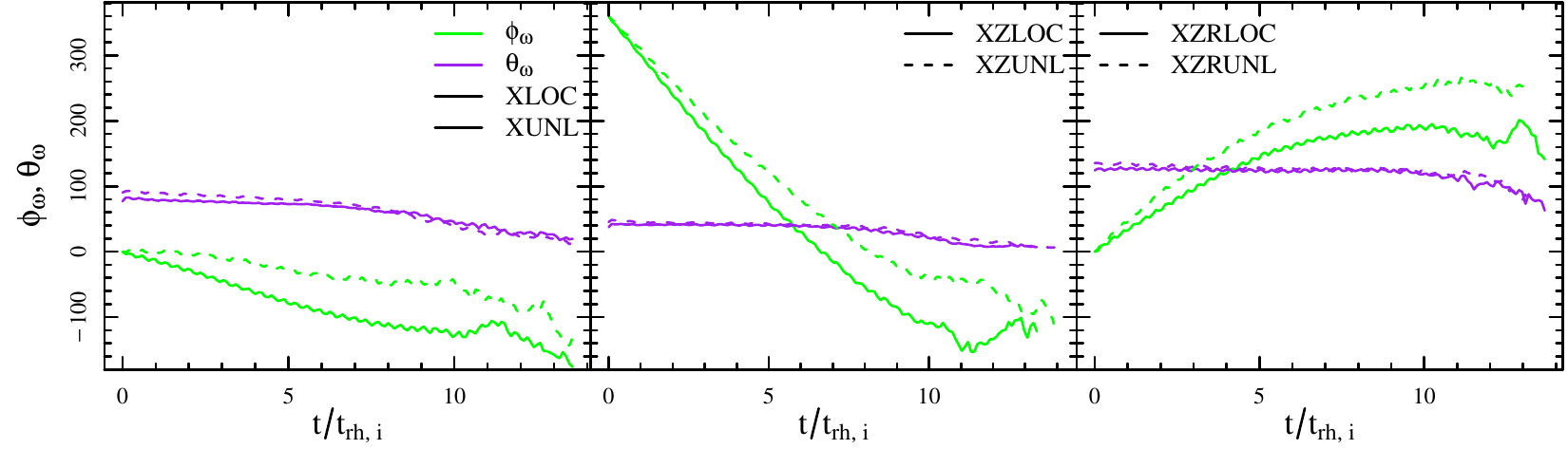}
    \caption{Time evolution of the three-dimensional orientation of the angular velocity vector, $\omegavec$, as expressed by $\phiomega$, i.e. the angle between the projection of $\omegavec$ on the $x-y$ plane and the $x$-axis, and $\thetaomega$, i.e. the angle between $\omegavec$ and the $z$-axis. All models are represented (see Table \ref{tab:table1}). All the systems are characterized by a precession and nutation of the angular velocity vector.}
    \label{fig:angleglob}
\end{figure*}

We begin our analysis by presenting the evolution of the components of the angular velocity vector, $\omegavec$, calculated using all particles within the Jacobi radius, $\rj$.  We calculate the moment of inertia tensor, $\mathbfss{I}$, and the angular momentum vector, $\mathbfit{L}$, and solve the equation $\mathbfit{L} = \mathbfss{I} \omegavec$.  The evolution of the absolute value $\omega$ and the Cartesian components 
($\omega_x$, $\omega_y$, $\omega_z$)  for all models are plotted in Fig.~\ref{fig:omglob} as a function of time, 
expressed in units of the initial half-mass relaxation time of the cluster, $\trhi$.

A number of interesting features are present in Fig. \ref{fig:omglob}. 
In agreement with previous findings 
(e.g., see \citealt{einsel1999,ernst2007,hong2013,tiongco2017}), the overall rotation in the system decreases in magnitude over time, 
as determined by a general redistribution and loss of angular momentum in the system due to relaxation effects.
As the system evolves, it gradually loses 
its intrinsic differential rotation and evolves towards a configuration dominated by an approximately solid-body rotation about the $z$-axis with angular speed equal to about 0.5$\Omega$ (models XZRLOC and XZRUNL take longer to converge to this value than the time shown in Fig. \ref{fig:omglob}). Thus, the rotating models explored here extend the conclusions of \citet{tiongco2016b}, namely that a
variety of initially non-rotating models reach a condition of only partial synchronization, as a result of the interaction with the external tidal field and the preferential loss of prograde orbiting stars.

A prominent feature in Fig.~\ref{fig:omglob} is that all models acquire a rotation about the $y$-axis. The initial rotation of the system in the x-z plane, the non-spherical symmetry
induced by this rotation, and its orientation imply that the system is affected by a tidal torque due to the host galaxy that initiates a precession of the rotation axis and a rotation about the $y$-axis (see the large scale oscillations in Fig. ~\ref{fig:omglob}). As the system loses its initial intrinsic rotation, the amplitudes of the oscillations of $\omega_y$ and $\omega_x$ are damped and eventually converge to zero leaving the system only with the rotation about the $z$-axis discussed above. The time evolution of $\omegavec$ presented in Fig. \ref{fig:omglob} also shows the presence of small scale and rapid nutation oscillations.

 We further explore the evolution of the angular velocity vector in terms of the angles ($\thetaomega$, $\phiomega$) which define the direction of $\omegavec$ (see Fig.~\ref{fig:angleglob}). 
  The evolution of $\phiomega$ illustrates the gradual 
change in orientation
 of  $\omegavec$ on the $x-y$ plane. Meanwhile, the evolution of $\thetaomega$ shows that, as the cluster loses its initial differential rotation about the $x$ and $y$ axis, the direction of $\omegavec$ gradually converges toward the $z$-axis. Both Figs.~\ref{fig:omglob} and \ref{fig:angleglob} indicate that all models are characterized by a precession and nutation oscillations of $\omegavec$. 

\begin{figure*}
	\includegraphics[height=2.6in]{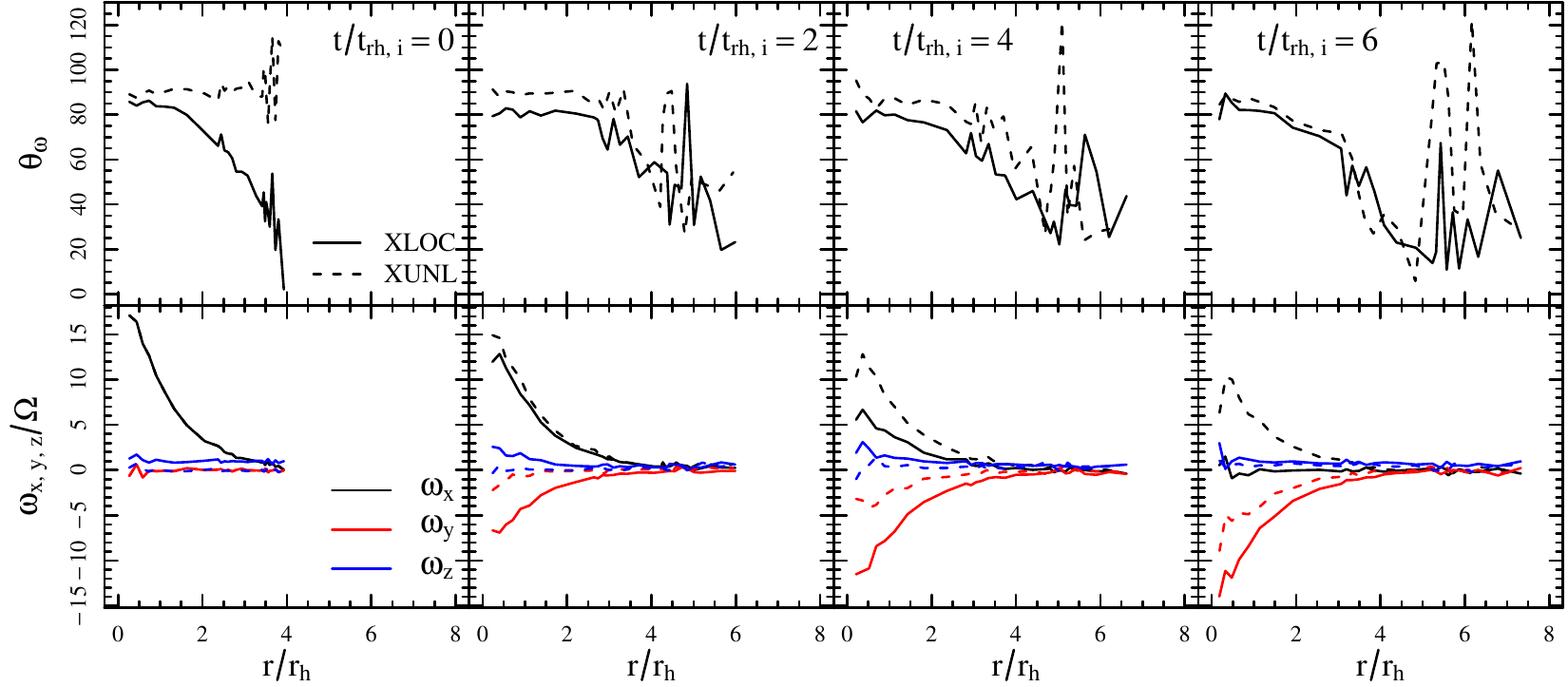}
    \caption{Top panels: Time evolution of the radial profile of $\thetaomega$ for models XLOC and XUNL.  Bottom panels: Evolution of the radial profiles of $\omega_x$, $\omega_y$, and $\omega_z$ of the same models. The radius is expressed in units of the half-mass radius of the cluster. As a result of the coupling with the orbital angular momentum, the orientation of the internal rotation axis of the models depends on the distance from the cluster's centre. }
    \label{fig:thetavsr}
\end{figure*}

We now turn our attention to 
the study of the rotational properties as a function of the distance from the cluster centre. The rotation curve of the \citet{varri2012} models is characterized by a radial profile increasing with radius in the cluster's inner regions, reaching a peak in the intermediate parts, and then decreasing and vanishing at the edge of the system. The additional solid-body rotation about the $z$-axis (either added in the initial conditions as in the locked systems or developing during the cluster evolution for the initially unlocked systems) naturally introduces a radial variation in the rotation axis.
We explore this 
feature by inspecting the time evolution of the radial profile of
$\thetaomega$ for models XLOC and XUNL (see Fig.~\ref{fig:thetavsr}, top row).
For model XLOC, a radial gradient is present in the initial conditions because of the added solid body rotation around the $z$-axis, and subsequently, the system is characterized by a radial gradient in the orientation of its rotation axis during its entire evolution.
Model XUNL, on the other hand, starts its evolution with no radial gradient in $\thetaomega$, but as it evolves and interacts with the external tidal field, it develops a radial profile of $\thetaomega$ similar to that of the XLOC model.

The radial profile of $\thetaomega$ shows that in the cluster's  innermost regions the orientation of the rotation axis is determined mainly by that of the cluster initial intrinsic rotation (in this case parallel to the $x$-axis); $\thetaomega$ decreases as the distance from the cluster center increases until the outermost regions where the rotation axis is almost parallel to the $z$-axis (i.e. parallel to the cluster's orbital angular velocity around the host galaxy) and mainly determined by the rotation resulting from the interaction of the cluster with the external tidal field. 

The bottom panels of Fig.~\ref{fig:thetavsr} show the radial profiles of $\omega_x$, $\omega_y$, and $\omega_z$, and further illustrate the radial variation of the cluster's rotational properties and the transition from the inner regions dominated by rotation around the $x$ and $y$ axes, to the outer regions that are most affected by the influence of the tidal field and are increasingly dominated by rotation around the $z$ axis.

\begin{figure*}
	\includegraphics[height=4.4in]{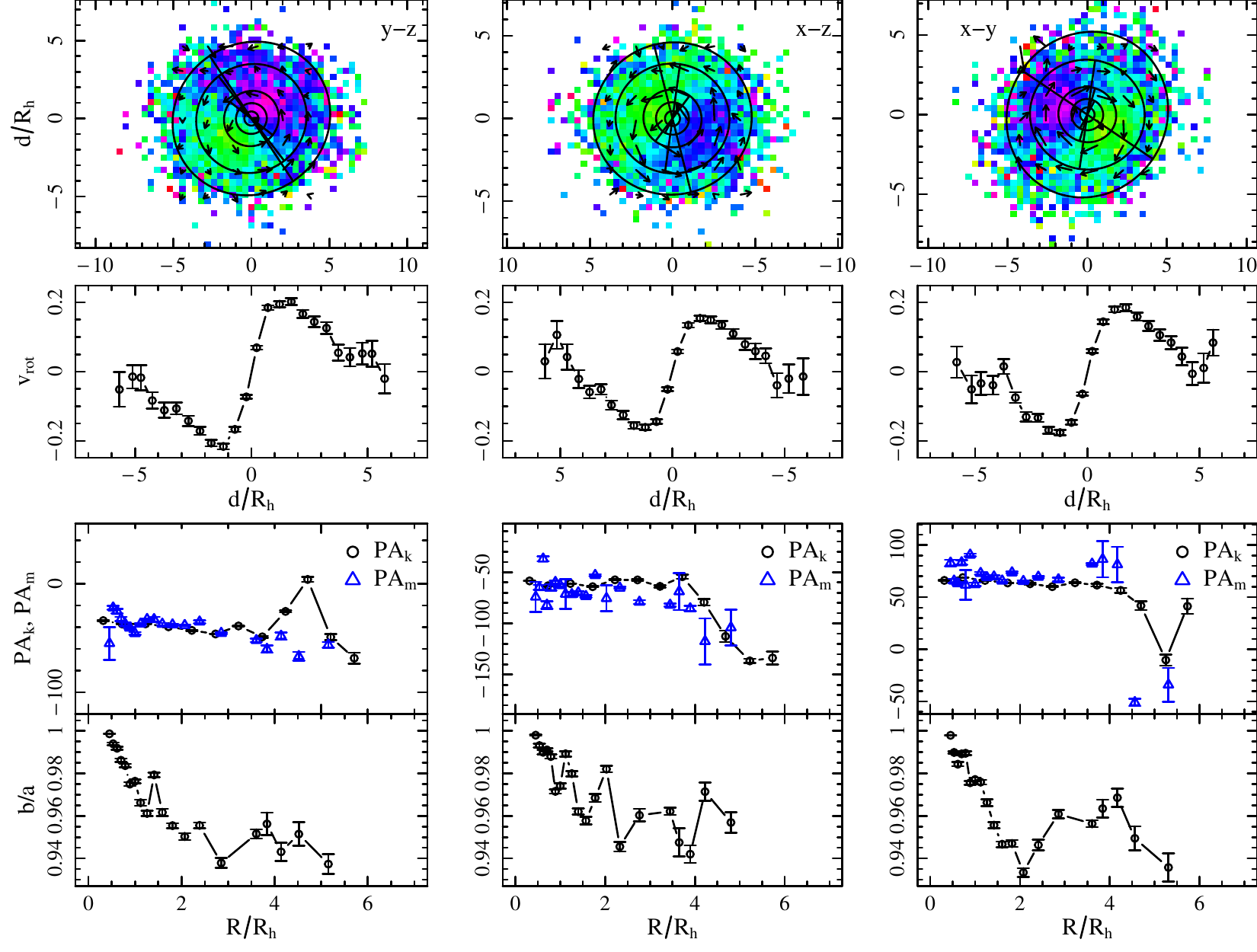}
    \caption{Analysis of a snapshot of the model XZRLOC at $t/\trhi=1.95$.  Each column presents a different projection of the cluster in the plane specified in the top panel.  First row: Cluster map colour coded according to the mean radial velocity of stars  in that area of the projection (yellow-green colors represent the lowest velocity, violet is the highest velocity), and superimposed arrows represent the mean tangential velocity of stars in that area of the projection. Each ellipse is an approximation of the isodensity contours, and the straight lines denote the minor axes.  Second row:  Rotation curve calculated in bins parallel to the position angle of the global rotation axis. The error bars represent the standard error of the mean velocity in each bin. Third row: Position angles of the projected rotation axes, $PA_{\rm k}$, and position angles of the projected minor axes, $PA_{\rm m}$ expressed as functions of the projected radius.  The error bars of $PA_{\rm k}$ are from the error of the fitting procedure.  Fourth row: Ratio of the projected minor to major axes versus the projected radius in units of the corresponding projected half-mass radius.  The error bars of $PA_{\rm m}$ and $b/a$ are from the standard error of the mean from bootstrap statistics of the positions of the stars in each bin.}
    \label{fig:veliso}
\end{figure*}

Although a detailed connection between our theoretical results and observations is beyond the scope of this paper, it is important to understand how and to what extent the kinematical features found in our simulations would appear when studied using the tools and diagnostics typically available in current observational studies.
To explore this issue, in Fig.~\ref{fig:veliso}, we concentrate on a representative snapshot of the model XZRLOC. 

The top panels of Fig.~\ref{fig:veliso} show colour maps representing the mean radial velocity along the line of sight adopted in each column; the arrows indicate the direction and relative magnitude of the velocity in the plane perpendicular to the line of sight.  
A rotation signature can be detected both in the radial velocity maps and in the tangential velocity vector fields.  This may simply suggest that the three-dimensional rotation axis is inclined relative to the plane of the projection, but as we have seen in Fig. \ref{fig:thetavsr}, the kinematical properties of our systems are in general more complex, and the direction of the rotation axis changes with the distance from the cluster centre.

The ellipses shown in the top panels are meant to be an approximation of projected isodensity contours.  They have been calculated by
diagonalizing the moment of inertia tensor of particles within elliptical annuli. The minor and major axes of such annuli are the square roots of the inverse of the first and second eigenvalue, respectively, and they are scaled to the size of the annulus. The orientation of the ellipse is calculated from the eigenvectors (a more detailed outline of the method can be found in \citealt{zemp2011}; here we use their method S1 adapted to two dimensions). The lines representing the minor axes show evidence of isodensity twisting. This feature is the morphological manifestation on the plane of projection of the  radial variation of the rotation axis from the cluster's inner regions, where rotation is dominated by the remaining initial intrinsic rotation, to the outer regions, that are most affected by the tidal field and are dominated by rotation around the $z$-axis.

  The second row shows a rotation curve calculated following the standard method adopted in many observational investigations to study the rotational properties from line-of-sight velocities (see e.g. \citealt{bellazzini2012}). Studies based on this method typically identify the orientation of a global rotation axis that maximizes the rotation signal, and either derive a single global value for the rotation amplitude or derive a rotation curve around that axis. The rotation present in our model is clearly evident in all projections.
We wish to point out that the radial variation of the rotation axis might in some cases manifest itself as a counter-rotation of the outermost regions relative to the inner ones (see \citealt{boberg2017} for a possible observational evidence of this effect). We stress, however, that a radial variation of the orientation of the rotation axis does not necessarily imply the presence of counter-rotation.

The third row shows the radial variation of the position angle of the rotation axis calculated following the method used for the rotation curves in the second row, but for individual circular annuli at different distances from the cluster centre.  
We note that, while in some projections the gradient is clearly visible, the strong intrinsic gradient may be significantly weaker in the projected kinematical data.
In the same panels, we also plot the radial variation of the position angle of the minor axes of the ellipses approximating the isodensity contours, and show that in general they follow the kinematical position angle. It is important to remark here that the radial variation in the orientation of the rotation axis and the cluster's minor axis may appear  misaligned  if the measurements of the cluster morphological and kinematical properties are made at the different distances from a cluster's centre.

Finally, in the fourth row of Fig. \ref{fig:veliso}, we show the radial variation  of the minor-to-major axis ratio, $b/a$, of the isodensity ellipses shown in the top row of the figure; these plots illustrate the effect of rotation and  external tidal field on the projected morphological properties as the distance from the cluster centre increases.

\section{Conclusions}

In this Letter, we have broadened the exploration of the effects of the interplay between the internal angular momentum and the external tidal field on the dynamics of collisional stellar systems.

While previous studies have explored the evolution of rotating clusters  with an initial internal angular velocity vector aligned with the orbital angular velocity one, here we have studied more general configurations with initial internal rotation axes pointing in generic directions, differing from that of the external orbital rotation.

Our simulations show that such a generalization of the cluster's initial conditions has a number of major implications and leads to a variety of complex features in the evolution of the cluster's rotational properties. Specifically:

1) We have followed the evolution of the total angular velocity vector, $\omegavec$, and shown that for a system with a rotation axis initially not aligned with the cluster's orbital angular velocity, the tidal torque due to the host galaxy initiates a precession and nutation of the cluster's rotation axis leading to a time-varying orientation and amplitude of the three components of $\omegavec$ (Figs. \ref{fig:omglob} and \ref{fig:angleglob}).

2) The effects of the external tidal field lead to an internal rotation about the z-axis (the axis parallel to the cluster's orbital angular velocity); this effect naturally results in a dependence of the orientation of the internal rotation axis on the distance from the cluster centre (Fig. \ref{fig:thetavsr}). The inner regions are dominated by the cluster intrinsic rotation with an orientation determined by the initial conditions and then evolving due to the effects of the axis precession mentioned above; the outer regions are dominated by an approximately solid-body rotation about the z-axis.

3) Depending on the direction of the initial intrinsic rotation, the radial variation of the rotation axis may lead to systems in which the outermost regions are in counter-rotation relative to the inner ones.

4) As the overall magnitude of rotation decreases with time due to redistribution and loss of angular momentum from the system, a cluster gradually loses its initial intrinsic rotation and is increasingly dominated by the rotation about the z-axis due to the effects of the external tidal field. The rotation axis precession mentioned above is therefore affected by the effects of internal relaxation leading to a gradual damping of the precession oscillations and a gradual convergence of the rotation axis toward the axis perpendicular to the cluster orbital plane ($z$-axis).

5) The radial variation of the orientation of the rotation axis (including its possible manifestation as counter-rotation) may be detected with radial velocity measurements, but projection effects play a significant role (Fig. \ref{fig:veliso}). Projected density maps show evidence of the radial variation of the cluster ellipticity and may also be characterized by a twisting in the isodensity contours.

The rich kinematics found in our simulations is the natural outcome of the evolution of rotating clusters starting from a general set of initial conditions. Such a degree of generality suggests that these properties might be widespread in globular clusters (with the exception of the dynamically oldest clusters that may have completely lost any memory of their initial rotation).

Our study provides guidance in the identification of non-trivial features which may emerge in upcoming investigations of star cluster kinematics and a dynamical framework to understand some of the complexities already hinted by recent observational studies.




\bibliographystyle{mnras}
\bibliography{references}






\bsp	
\label{lastpage}
\end{document}